\documentclass{desyproc}
\begin{document}
\title{Diffraction dissociation at the LHC}

\author{{\slshape L\'aszl\'o Jenkovszky$^{1,2}$, Andrii Salii$^1$}\\[1ex]
$^1$Bogolyubov Institute for Theoretical Physics (BITP), Ukrainian National Academy of Sciences \\14-b, Metrolohichna str., Kiev, 03680, Ukraine\\
$^2$Wigner Research Centre for Physics, Hungarian Academy of Sciences\\
1525 Budapest, POB 49, Hungary}


\maketitle

\begin{abstract}
Predictions for the squared momentum transfer and missing mass dependence of the differential and integrated single- and double low missing-mass diffraction dissociation in the kinematical range of present and future LHC measurements are summarized.
\end{abstract}
In a recent series of papers \cite{PR,2,3} a dual-Regge model for high-energy low-missing mass diffraction dissociation was developed. The model is based on Regge factorization,
exact in a the case of a single Regge-pole exchange. As argued in Ref. \cite{JLL}, low-$t$ data at the LHC are dominated by a single Pomeron exchange, hence factorization at the LHC is applicable there. Regge factorization relates elastic scattering to single- and double diffraction dissociation.
Really, by writing the scattering amplitude as product of the vertices, elastic $f$ and inelastic $F$, multiplied by the (universal) propagator (Pomeron exchange), $f^2s^{\alpha} \ \ fFs^{\alpha},\ \ F^2s^{\alpha}$ for elastic scattering, SD and DD, respectively, one gets
 {\footnotesize
\begin{equation}\label{factor1}
\frac{d^3\sigma_{DD}}{dtdM_1^2dM_2^2}=\frac{d^2\sigma_{SD1}}{dtdM_1^2}\frac{d^2\sigma_{SD2}}{dtdM_2^2}/\frac{d\sigma_{el}}{dt}.
\end{equation}}
\begin{figure}[!ht]
\centerline{\includegraphics[width=0.7\textwidth,bb=0 420 945 627,clip]{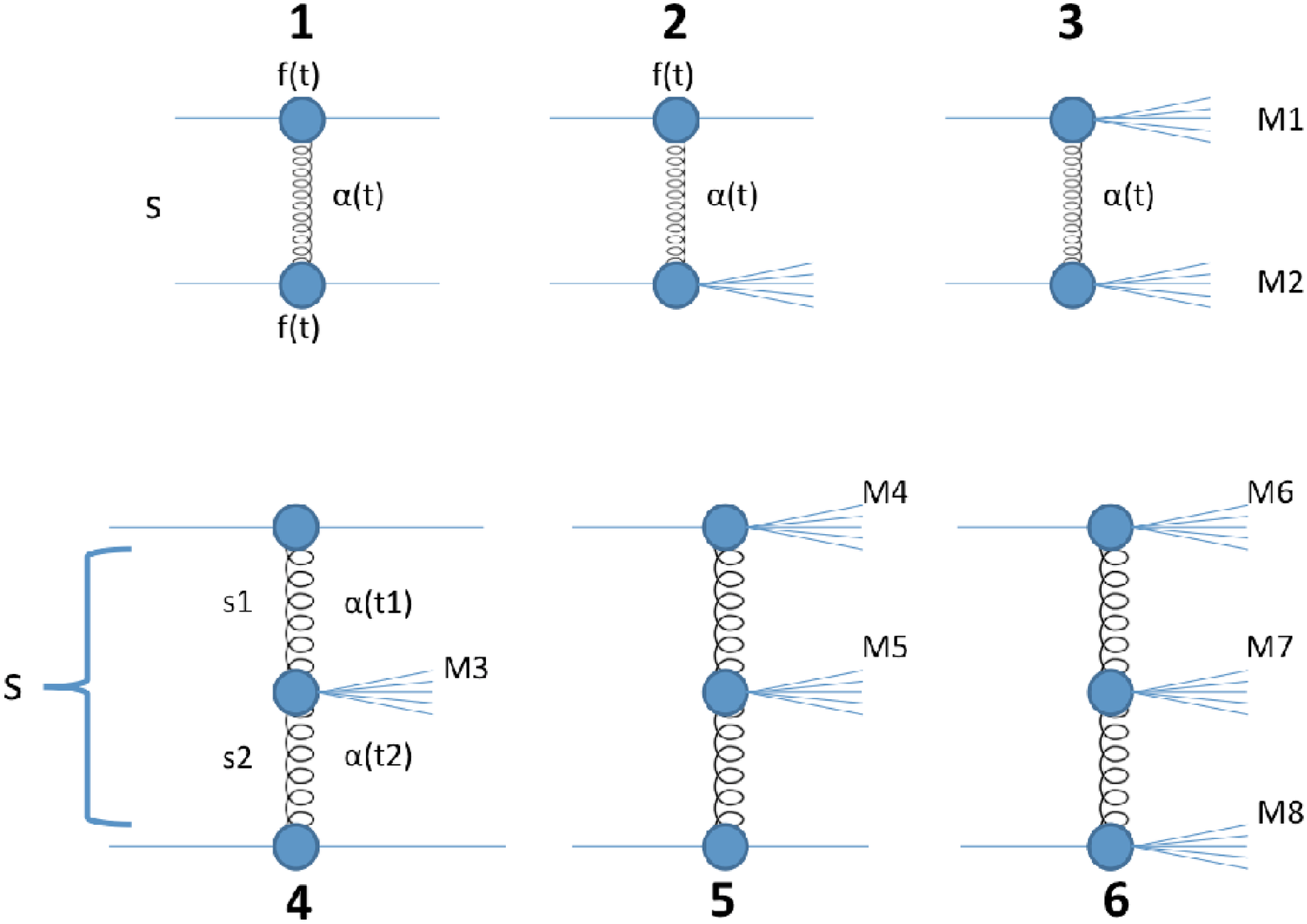}}
\centerline{\includegraphics[width=0.7\textwidth,bb=0 18 950 350,clip]{DD.eps}}
\caption{Diagrams for elastic scattering and diffraction dissociation (single double and central).} \label{fig:FeynmanDiagrams}
 \label{fig:Pp}
\end{figure}

Another important ingredient of the present model is the inelastic $pPX$ vertex, or $pP$ total cross section entering the diagrams of Fig. {\ref{fig:Pp}} , where $P$ stands for the Pomeron and
$X$ is the the diffractively produced state (proton's excitation). Following Ref. \cite{JL}, we assume that the Pomeron is similar to the virtual photon and the inelastic vertex (transition amplitude)
has the same properties as the structure function in deep inelastic photon-nucleon scattering, known e.g. from JLab or HERA experiments. We also make full use of duality, incorporating both resonance (in the missing mass $M$) and Regge asymptotic, for high missing masses, of the $Pp$ transition amplitude. The direct channel, i.e. in $M$ is dominated by the known non-linear nucleon trajectory. For more details see Refs. \cite{PR,3}.

Below we sumamrize some of our results on single- and double diffraction cross sections as functions of the incoming squared energy, $s$, squared momentum transfer $t$ as well as the missing masses $M$. 

 In the calculations the following formulae for elastic, SD and DD  scattering were used:
  {\footnotesize
  \begin{equation}
  \label{elastic+SD}\frac{d\sigma_{el}}{dt}=A_{el}{F_p}^4(t)\left(\frac{s}{s_0}\right)_,^{2(\alpha(t)-1)}\;\;
   2\cdot\frac{d^2\sigma_{SD}}{dtdM_x^2}={F_p}^2(t){F_{inel}}^2(t,M_x^2) \left(\frac{s}{M_x^2}\right)_,^{2(\alpha(t)-1)}\end{equation}
  \begin{equation}\label{DD}\frac{d^3\sigma_{DD}}{dtdM_1^2dM_2^2}=N_{DD}{F_{inel}}^2(t,M_1^2){F_{inel}}^2(t,M_2^2)\left(\frac{ss_0}{M_1^2M_2^2}\right)_.^{2(\alpha(t)-1)}\end{equation}}
Where the norm: {\footnotesize$N_{DD}=\frac{1}{4A_{el}},$}
the inelastic vertex:
 {\footnotesize ${F_{inel}}^2(t,M_x^2)=A_{res}\frac{1}{M_x^4}\sigma_T^{Pp}(M_i^2,t)+C_{bg}\sigma_{Bg},$}\\
 the Pomeron-proton total cross section, which is the sum of $N^*$ resonances and the Roper resonance, with a relevant norm $R$ (see \cite{PR}):
 {\footnotesize
    $$\sigma_T^{Pp}(M_x^2,t)=
    R\frac{[f_{res}(t)]^{2} \cdot M_{Roper}\left(\frac{\Gamma_{Roper}}{2}\right)}
    {{\left(M_x^2-M_{Roper}^2\right)}^2+{\left(\frac{\Gamma_{Roper}}{2}\right)}^2}
    +[f_{res}(t)]^{4}\sum_{n=1,3} \frac{\Im m\,\alpha(M^2_x)}{(2n+0.5-{\Re}e\, \alpha(M_x^2))^2+({\Im}m\,\alpha(M^2_x))^2},$$}
and the background contribution:
 { $\sigma_{Bg}=\frac{f_{bg}(t)}{\frac{1}{{\left(M_x^2-(m_p+m_{\pi})^2\right)}^\varsigma}+(M_x^2)^{\eta}}.$}

 We use the Pomeron trajectory \cite{JLL}: {\footnotesize$\alpha(t)=1.075+0.34t$}, and the form factors:
  $F_p(t)=e^{b_{el}t},$ $f_{res}(t)=e^{b_{res}t},$ $f_{bg}(t)=e^{b_{bg}t}.$

The values of the fitted parameters are presented in Table \ref{tab:ParSet};
our prediction  are summarized in Table \ref{tab:cs.predict}.
The $\chi^2$ values are quoted in relevant figures.
\begin{table}[!ht]
\centering
  \begin{tabular}{ccc}
     \begin{tabular}{|c||c|}
        \hline
        $A_{el}[mb]$        & $33.579$\\
        $b_{el}[$GeV$^{-2}]$  &$1.937$\\ \hline

    $A_{res}[mb\cdot$GeV$^4]$&$2.21$\\
        $C_{bg}[mb]$        &$2.07$\\
        $R$                 &$0.45$\\
        $b_{res}[$GeV$^{-2}]$ &$-0.507$\\
        $b_{bg}[$GeV$^{-2}]$  &$-1.013$\\ \hline
     \end{tabular}&\qquad\qquad&
      \begin{tabular}{|c||c|}
        \hline
        $s_0$   &$1$\\ \hline
        $\varsigma$ &$0.8$\\
        $\eta$      &$1$\\ \hline
        \hline
        \multicolumn{2}{|c|}{$\alpha(t)=\alpha(0)+\alpha't$}\\ \hline
        $\alpha(0)$           & $1.075$\\
        $\alpha'$[GeV$^{-2}$] & $0.34$\\\hline
     \end{tabular}%
  \end{tabular}%
 \caption{\label{tab:ParSet} Fitted parameters, see Eqs.~(\ref{elastic+SD}), (\ref{DD}).}
\end{table}

\begin{figure}[!ht]
 \centerline{
  \includegraphics[width=0.49\linewidth,bb=13mm 24mm 195mm 185mm,clip]{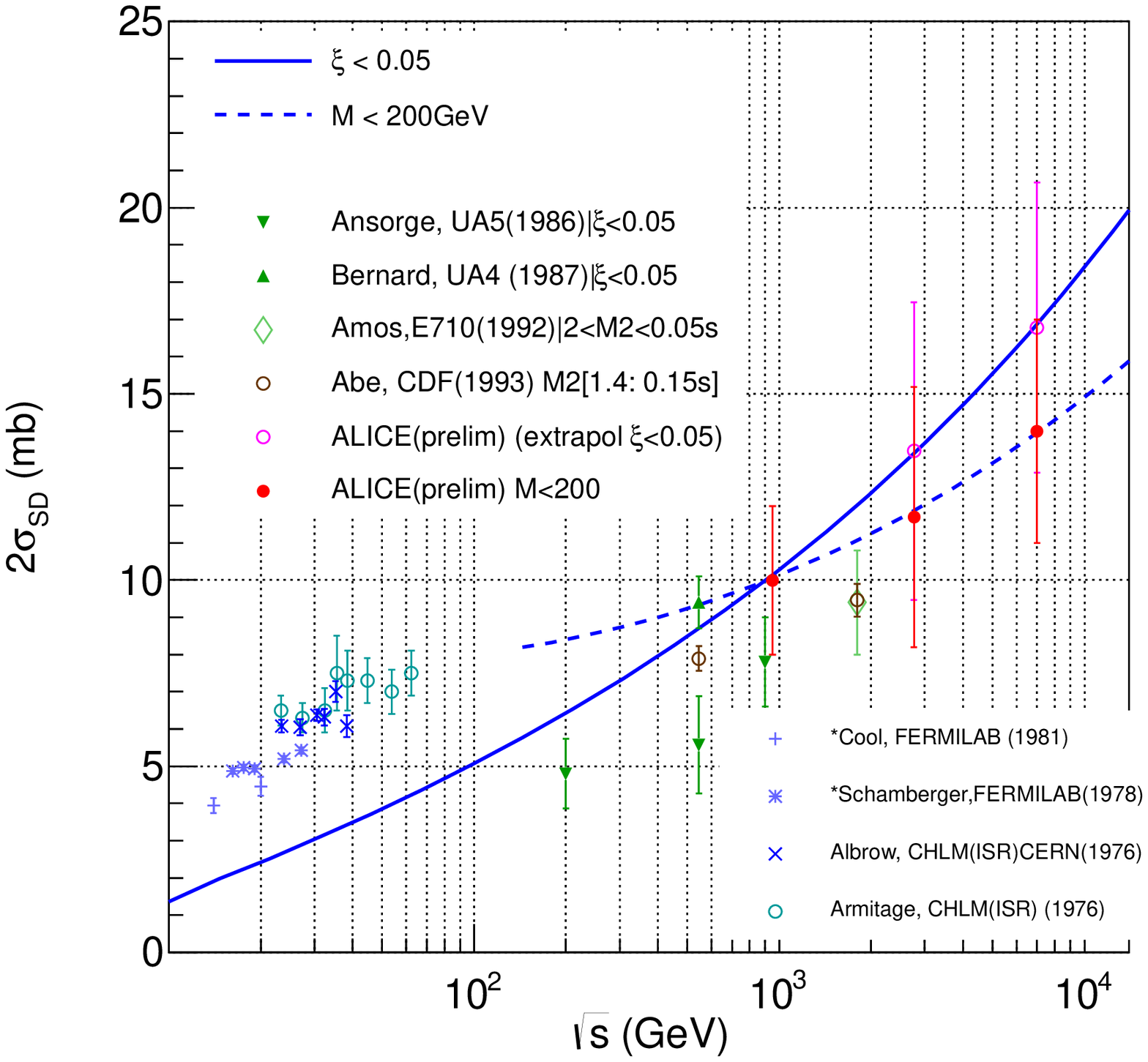}
  \includegraphics[width=0.49\linewidth,bb=13mm 24mm 195mm 185mm,clip]{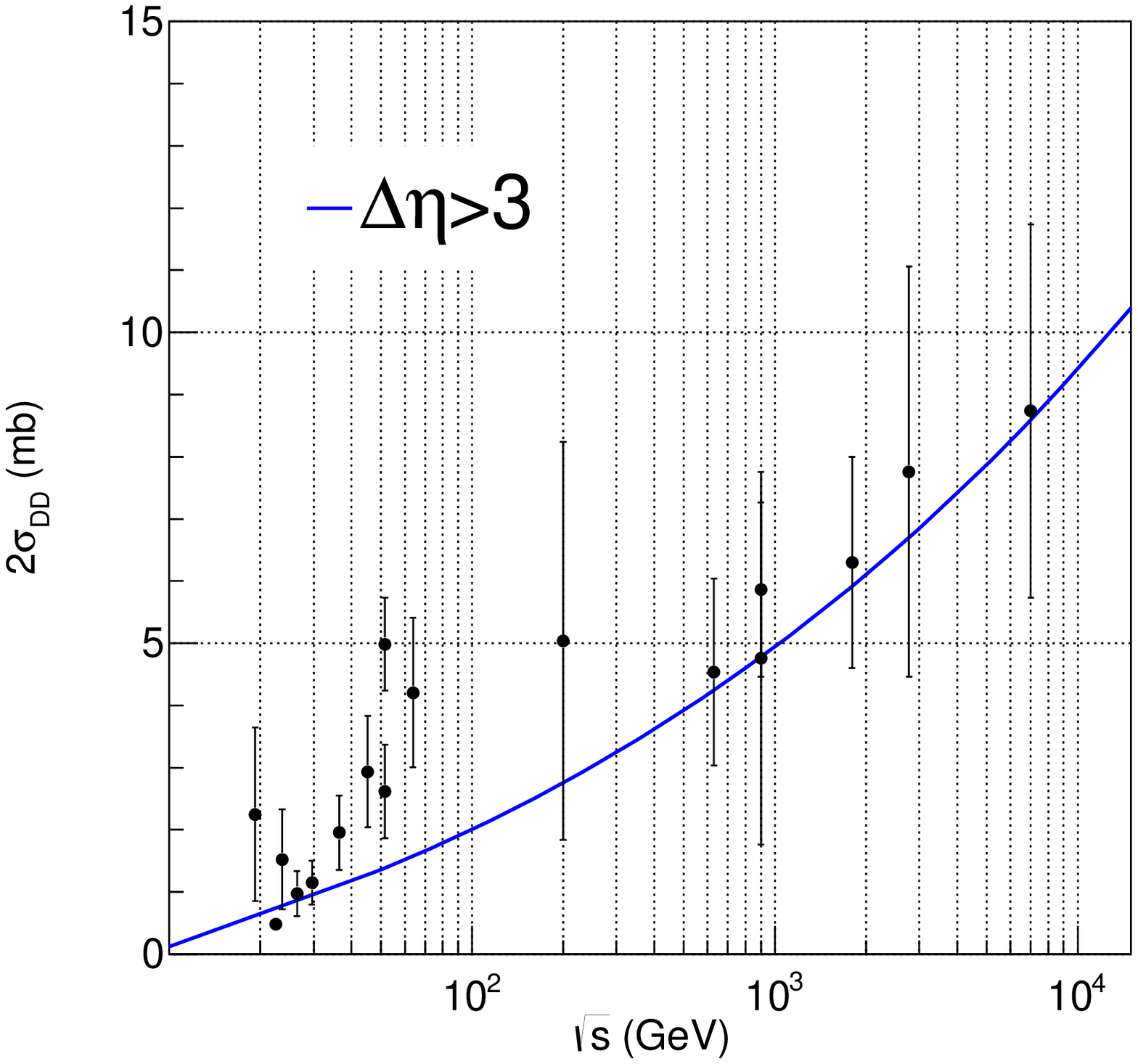}}
  \caption{(a) Single diffraction dissociation cross section vs. energy $\sqrt{s}$ calculated from Eq. (\ref{elastic+SD}). The points with red full circles correspond to (unpublished) ALICE data ($M<200GeV$), the point with pink open circles are the same points extrapolated to $\xi<0.05$. For $\sigma|_{\xi<0.05}$: $\chi^2/n=1.6$, $n=6$ ($\sqrt{s}>100GeV$ only); for $\sigma|_{M<200GeV}$: $\chi^2/n<0.01$, $n=3$.
  (b) Double diffraction dissociation cross section vs. energy $\sqrt{s}$ calculated from Eq. (\ref{DD}). $\chi^2/n=0.2$, $n=7$, for $\sqrt{s}>100GeV$ only.
 Experimental data are from \cite{Poghosyan for ALICE}.}
  \label{fig:cs.SD|cs.DD.Data}
\end{figure}
\begin{figure}[!hb]
 \centerline{\includegraphics[width=0.38\linewidth,bb=13mm 10mm 195mm 185mm,clip]{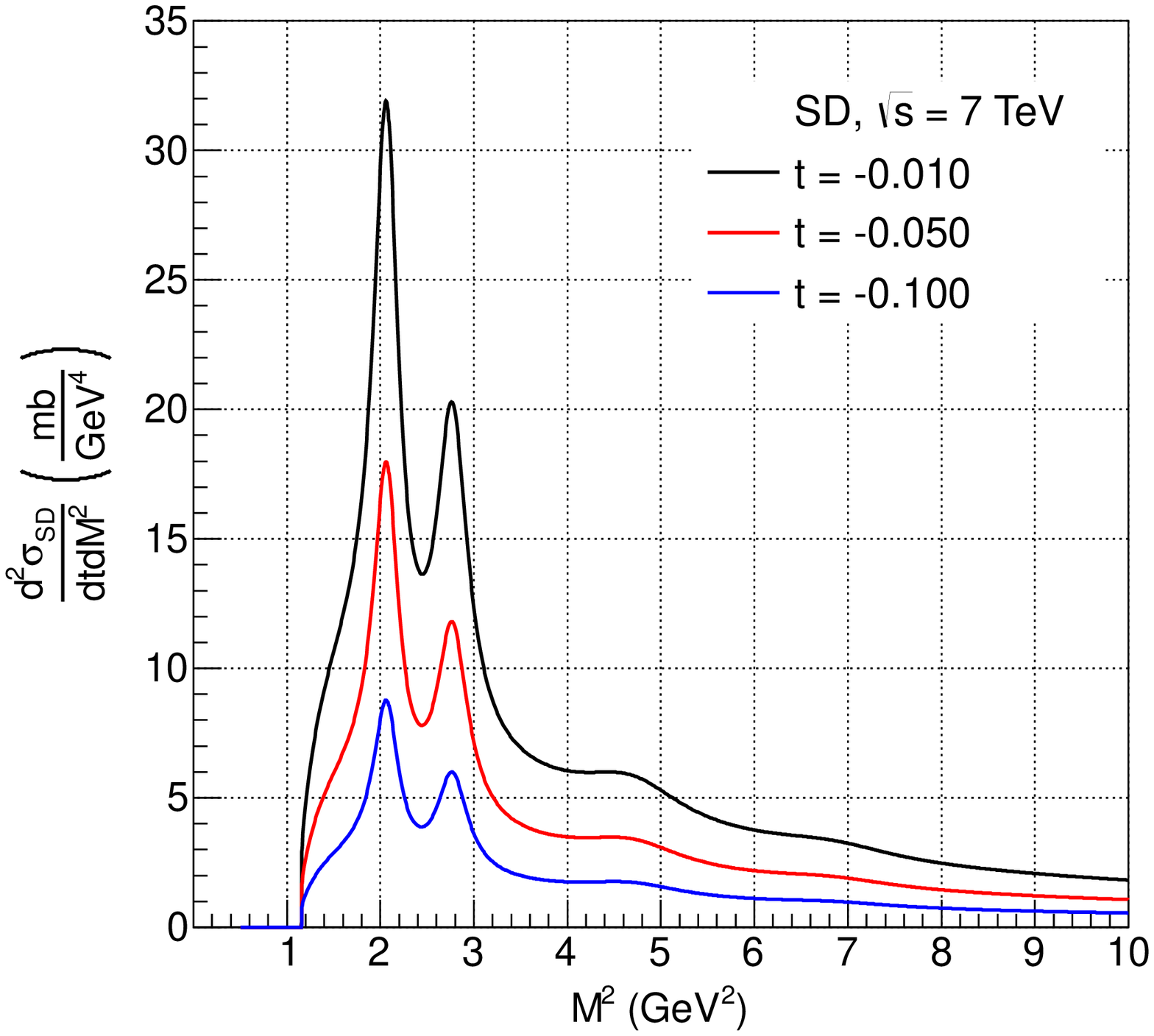}
 \includegraphics[width=0.60\linewidth,bb=0mm 0mm 490mm 305mm,clip]{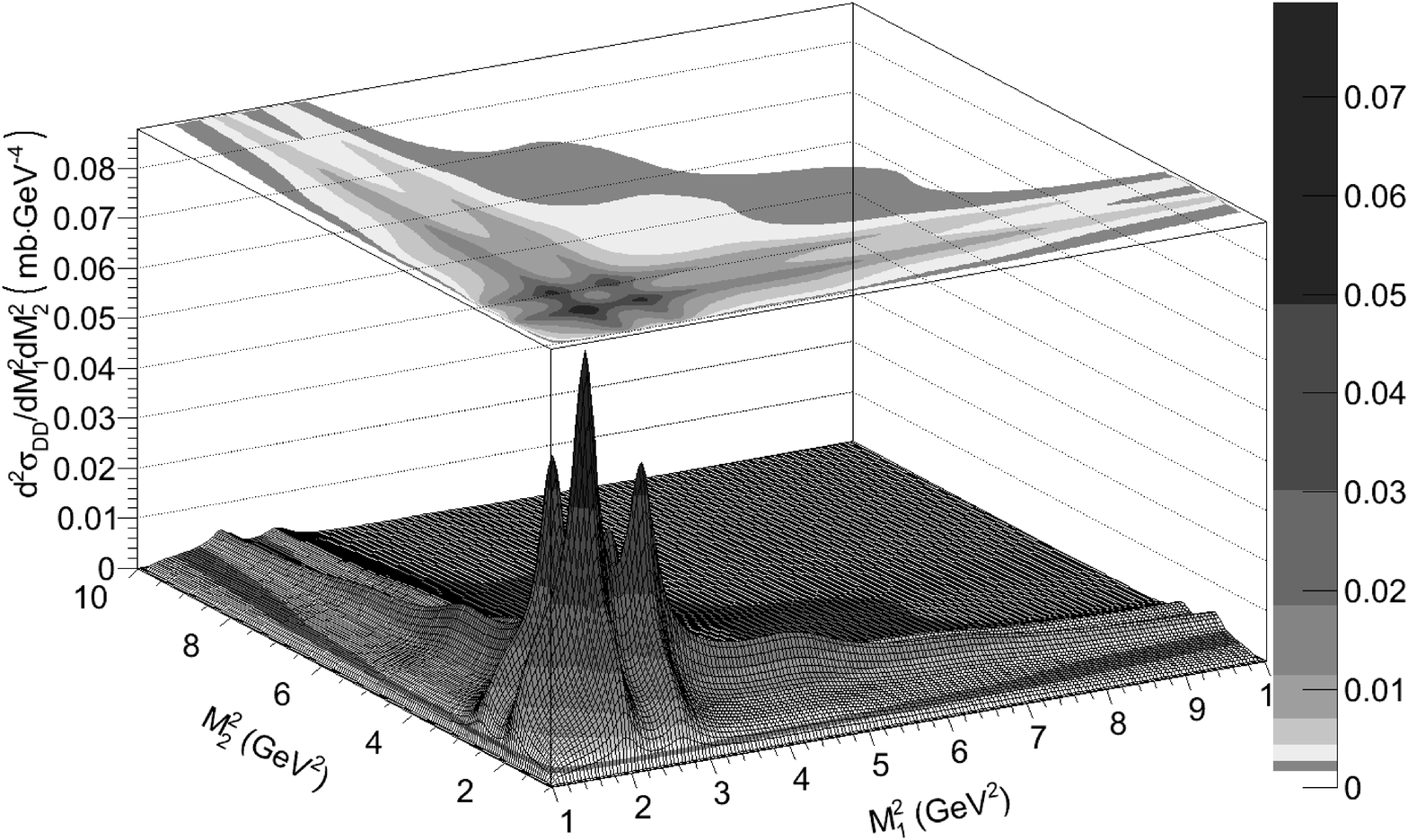}}
 \caption{(a) Double differential SD cross sections as functions of $M^2$ for different $t$ values. \quad
   (b) Double differential DD cross sections as functions of $M_1^2$ and $M_2^2$ integrated over $t$.}
  \label{d2cs_m2.in.t}
\end{figure}
\section*{Conclusions}\label{sec:Conclusions}
At the LHC, in the diffraction cone region ($t<1$ GeV$^2$) proton-proton scattering is dominated by Pomeron exchange (quantified in Ref. \cite{JLL}). This enables full use of factorized Regge-pole models. Contributions from non-leading (secondary) trajectories can (and should be) included in the extension of the model to low energies, e.g. below those of the SPS.

%
Unlike most of the approaches that use the triple Regge limit in calculating diffraction dissociation, our approach is based on the assumed similarity between the Pomeron-proton and virtual photon-proton scattering. The proton structure function (SF) probed by the Pomeron is the central object of our studies. This SF, similar to the DIS SF, is exhibits direct-channel (i.e. missing mass, $M$) resonances transformed in resonances in single- double- and central diffraction dissociation. The high-$M$ behaviour of the SF (or Pomeron-proton cross section) is Regge-behaved and contains two components, one decreasing roughly as $M^{-m},\ \ m\approx 2$ due to the exchange of a secondary Reggeon.


The results of our calculations, that are mainly predictions for the LHC energies $7,\ 10$ and $14$ TeV, are collected in Table \ref{tab:cs.predict}.The quality of the fit is quantified by the relevant $\chi^2$ values.

\begin{wraptable}{r}{0.45\linewidth}
  \centerline{ \begin{tabular}{|c||c|c||c|}
       \hline
       $\sqrt{s}$& $\sigma_{SD}$ & $\sigma_{SD}$& $\sigma_{DD}$\\
        {\footnotesize$[TeV]$}  &{\footnotesize${M^2<200GeV}$}&{\footnotesize ${\xi<0.05}$}&{\footnotesize${\Delta\eta>3}$}\\  \hline
       7 &13.96& 16.87& 8.63 \\
       8 &14.30& 17.43& 8.93 \\
       13&15.67& 19.59&10.09 \\ \hline
   \end{tabular}}
   \caption{\label{tab:cs.predict}Predictions for the LHC (in mb).}
\end{wraptable}

Our approach is inclusive, ignoring e.g. the angular distribution of the produced particles from decaying resonances.
All resonances, except Roper, lie on the $N^*$ trajectory. Any complete study of the final states should included also spin degrees of freedom, ignored in the present model.

%
For simplicity we used linear Regge trajectories and exponential residue functions, thus limiting the applicability of our model to low and intermediate values of $|t|$. Its extension to larger $|t|$ is straightforward and promising. It may reveal new phenomena, such as the the possible dip-bump structure is SD and DD as well as the transition to hard scattering at large momenta transfers, although it should be remembered that
diffraction (coherence) is limited (independently) both by $t$ and $\xi$.

\section*{Acknowledgments}
We thank Oleg Kuprash, Vladimir Magas and Risto Orava for their collaboration on the 
subject of this presentation. 
L.J. thanks the Organizers of this Conference for their hospitality.
This work is partially supported by WP8 of the hadron physics program of
the 8th EU program period.

\begin{footnotesize}

\end{footnotesize}
\end{document}